\documentclass[10pt]{IEEEtran}
\usepackage{epsfig,endnotes,url,color}
\usepackage{array}
\usepackage[english]{babel}
\usepackage{xspace}
\usepackage{verbatim}
\usepackage[ruled, vlined]{algorithm2e}
\usepackage{amsmath}
\usepackage{caption}
\usepackage{fancyhdr}
\usepackage[caption=false,font=footnotesize]{subfig}
\usepackage{balance}

\usepackage{mathpazo} 
\newcommand{\myparagraph}[1]{\smallskip \noindent{\bf {#1}.}}

\newcommand{\out}[1] {}

\newcounter{codeLineCntr}

\reversemarginpar
\newif\ifnotes
\notestrue

\newcommand{\punt}[1]{}

\renewcommand{\eqref}[1]{Equation~(\ref{eq:#1})}

\newcommand{\proc}[1]{\ifmmode\mbox{\textsc{#1}}\else\textsc{#1}\fi}

  \newcommand{\func}[1]{\ifmmode\mathrm{#1}\else\textrm{#1}fi} %

\newcounter{remark}[section]

\newcommand{\myfontsize}{\fontsize{7}{9}\selectfont}

\newcommand{\projecttitle}{\textsc{Inspector}\xspace}
\newcommand{\pthreads}{{\tt pthreads}\xspace}

\newcommand{\intelpt}{{Intel PT}\xspace}

\begin{document}

\author{
\IEEEauthorblockN{J{\"o}rg Thalheim, Pramod Bhatotia, and Christof Fetzer}\\
\IEEEauthorblockA{TU Dresden}\\
}

\title{\Large \projecttitle: A Data Provenance Library for Multithreaded Programs}
\maketitle
\begin{abstract}
Data provenance strives for {\em explaining} how the computation was performed by recording a trace of the execution. The provenance trace is useful across a wide-range of workflows to improve the dependability, security, and efficiency of software systems. 

In this paper, we present \projecttitle, a {\tt POSIX}-compliant data provenance library for shared-memory multithreaded programs. The \projecttitle library is completely transparent and easy to use: it can be used as a replacement for the \pthreads library by a simple exchange of libraries linked, without even recompiling the application code.

To achieve this result,  we present a parallel provenance algorithm that records control, data, and schedule dependencies using  a {\em Concurrent Provenance Graph} (CPG).  We implemented our algorithm to operate at the compiled binary code
level by leveraging a combination of OS-specific mechanisms, and recently released \intelpt ISA extensions as part of the Broadwell micro-architecture.  Our
evaluation on a multicore platform using applications from multithreaded benchmarks suites (PARSEC
and Phoenix) shows reasonable provenance overheads for a majority of applications.

Lastly, we briefly describe three case-studies where the generic interface exported by \projecttitle is being used to improve the dependability, security, and efficiency of systems. The \projecttitle library is  publicly available for further use in a wide range of other provenance workflows.

\end{abstract}

\section{Introduction}
\label{sec:introduction}

A data provenance-aware system gathers and reports the lineage of execution. This allows the user to track, and understand, how the computation was performed.  The provenance trace is useful for a wide-range of workflows to improve the dependability, security, and efficiency of software systems; including, program debugging~\cite{fast-track-pldi}, state machine replication~\cite{rex},  compiler optimizations~\cite{pgo}, incremental computation~\cite{ithreads}, program slicing~\cite{roly}, memory management~\cite{memprof}, and dynamic information flow tracking~\cite{dift}, etc.

More specifically, the data provenance trace provides an explicit intermediate program representation recording control and data dependencies for a program execution. 
Many existing systems provide support for data provenance (details in $\S$\ref{sec:related}); however,
most existing solutions target sequential programs (or at the granularity of the entire process), while others that do support parallelism rely on restrictive application-specific programming model. As a result, the existing solutions have limited adoption in practice for the general shared-memory multithreaded programs.

In this paper, we propose an operating systems-based approach to data provenance for multithreaded programs. More specifically, we have the following three main design goals: 
\begin{itemize} 

\item Transparency: To support unmodified multithreaded programs without requiring any code changes to existing applications. 
\item Generality: To support the general shared-memory programming model with the  full range of synchronization primitives in the {\tt POSIX} API. 
\item Efficiency: To impose low overheads by designing the underlying provenance algorithm to be  {\em parallel} as well so that it does not limit the available application parallelism.

\end{itemize}

To achieve these goals, we present \projecttitle, a data provenance library for multithreaded programs. We implemented \projecttitle as a dynamically linkable shared library. To run a program using \projecttitle,  the user just needs to preload the \projecttitle library, and then, run the program as usual. Thus, our library supports existing binaries without any code changes or re-compilation. The library exports the provenance information to the {\tt perf} utility as an extended interface.

To run a program using \projecttitle,  the user just needs to preload the \projecttitle library  by using the environment variable {\tt LD\_PRELOAD} or {\tt -rdynamic} flag, and then, run the program as usual.

Our high level approach is based on recording data, control, and schedule dependencies in a computation by constructing a {\em Concurrent Provenance Graph} (CPG). The CPG tracks the input data to a program, all sub-computations (a sub-computation is a unit of the computation), the data flow between sub-computations, intra-thread control flow, and inter-thread schedule dependencies for the multithreaded execution.

In this paper, we present a {\em parallel} algorithm to build the CPG. Our algorithm leverages the Release Consistency (RC) memory model~\cite{DSM-RC} to efficiently record the inter-thread data and schedule dependencies in a completely decentralized manner. We implemented our algorithm as a dynamically linkable shared library by leveraging process-level isolation, MMU-assisted memory tracking, and \intelpt ISA extensions, released recently as part of the Broadwell micro-architecture.  Furthermore, we extended the library to support a consistent snapshot facility, where the user can analyze the provenance on-the-fly while the program is still running.

In particular, we make the following contributions:
\begin{itemize}

\item We present a parallel algorithm for data provenance for multithreaded programs that records control, data, and schedule dependencies using a Concurrent Provenance Graph (CPG) ($\S$\ref{sec:algorithms}).

\item We implemented our algorithm as a dynamically linkable shared library, which we call \projecttitle, leveraging MMU-assisted memory tracking, process-level isolation, and \intelpt ISA extensions.  The \projecttitle library can be loaded and linked at run-time as a replacement to the \pthreads library, without any recompilation  of the application code ($\S$\ref{sec:implementation}).

\item We further extended the library to support a live snapshot facility, where the user can analyze the provenance on-the-fly while the program is still running. The library periodically takes a consistent snapshot~\cite{chandy-lamport} of the CPG in a decentralized fashion ($\S$\ref{sec:snapshot}).

\end{itemize}

We  empirically demonstrate  the effectiveness of \projecttitle by applying it to applications of  PARSEC~\cite{parsec} and Phoenix~\cite{phoenix} benchmark suites. Our experiments show that \projecttitle~incurs reasonable overhead to record data provenance for a majority of applications ($\S$\ref{sec:evaluation}). 

Furthermore, we briefly describe three on-going projects where the generic provenance interface exported by \projecttitle is being used to improve the dependability, security, and efficiency of software systems  ($\S$\ref{sec:discussion}). \projecttitle is an active open-source project and the library is publicly available to the research community for further use in other workflows.

\section{Overview}
\label{sec:overview}

Our approach targets a shared-memory multithreaded environment, in which threads parallelize computation and take 
advantage of shared portions of the address space to efficiently communicate with each other. Apart from performing reads 
and writes to the shared memory, threads also employ different types of synchronization mechanisms to 
co-ordinate their progress, thereby ensuring correct semantics. 

We base our design on {\tt POSIX} threads, commonly referred to as
{\tt pthreads}, which is a widely used threading library for shared-memory
multithreading with a rich set of synchronization primitives.  This
choice has several advantages, namely that the {\tt POSIX} interface
is standardized across different architectures and operating systems. Furthermore, {\tt pthreads} is used as the underlying threading
library for many higher level abstractions for parallel programming
(e.g., {\tt OpenMP}). Therefore, our design choice benefits a lot of existing applications.

\myparagraph{Basic approach} At a high level, we record data provenance for a multithreaded execution by constructing a {\em Concurrent Provenance Graph (or CPG)}. Informally,  the CPG records three types of dependencies; namely, control, data, and schedule dependencies for the multithreaded execution. To record these dependencies, we divide thread execution into sub-computations. We record the execution trace to construct the CPG that tracks the {\em data flow} between the sub-computations, {\em control flow} for each thread execution, and threads interleaving or {\em schedule dependency}  in the multithreaded execution.

More specifically, the Concurrent Provenance Graph (or CPG) records a partial order $O = (N, \rightarrow$) among sub-computations with the following property: given a sub-computation $n$ (where $n \in N $)  and the subset of sub-computations $M$ that precede it according to $\rightarrow$, i.e., $M = \{M \subset N \mid \forall m \in M,$ $m \rightarrow n\}$, if the writes made by $m$ becomes visible to $n$ then the partial order $\rightarrow$ captures this possible data flow between sub-computations.

\myparagraph{Example} Using a simple example (shown in Figure~\ref{fig:simple-example}), we next explain how we record these dependencies for a shared-memory multithreaded program. The example considers a multithreaded execution with two threads ($T_1$ and $T_2$) modifying two shared variables ($x$ and $y$) using a lock. In the example, we assume that a thread execution is divided into sub-computations at the boundaries of synchronization primitives, such as {\tt lock()/unlock()}. (We explain the reason behind this design choice in $\S$\ref{sec:model}.) We identify these sub-computations as $T_1.a$ and $T_1.b$ for thread $T_1$, and $T_2.a$ for thread $T_2$.   To understand the dependencies that need to be recorded for the required partial order ($\rightarrow$), we showcase three cases for recording the control, schedule, and data dependencies.

The first dependency that we need to record is the {\em control flow} execution of each thread. In particular, we need to record the intra-thread execution order of sub-computations. For example, sub-computation $T_1.b$ follows $T_1.a$, and therefore, the control flow dependency records this partial order as $T_1.a \rightarrow T_1.b$. Additionally, we need to record the control flow path taken within a sub-computation. For example, sub-computation  $T_1.a$ has a conditional branch ({\tt if/else}) based on the value of {\tt flag}. We supplement the control flow dependency with all control paths taken by a thread within each sub-computation; i.e.,  all branches taken at run-time.

Secondly, we need to record the inter-thread {\em schedule} dependencies. The sub-computations can be interleaved in different order across executions because of the non-deterministic thread scheduling by the underlying OS. For instance, when threads acquiring the lock in the reverse order where thread $T_1.b$ gets acquire the lock before $T_2.a$. In this case, the final value of $y$ is affected based on this new ordering. Therefore, we also need to record the schedule dependencies between sub-computations as part of the partial order. We record these schedule dependencies by tracking interleaving of sub-computations by recording the thread schedule (For example, $T_1.a   \rightarrow   T_2.a \rightarrow T_1.b $).

Lastly,   we need to record {\em data dependencies} between sub-computations as a part of the partial order. For that, we track read and write sets for each sub-computation, i.e., the set of memory locations read or written by the sub-computation, respectively. The data dependencies are recorded implicitly using read and write-sets, and the partial order recorded using the control and schedule dependencies: if we know what data is read and written by each sub-computation, we can determine whether a data dependency exists by following the partial order, i.e. if a sub-computation is {\em transitively} reading the data that was modified by a sub-computation that precedes it in the partial order $\rightarrow$ then there exists a read-after-write data dependency.

\begin{figure}[t]
\centering
\myfontsize
{
\begin{tabular}
{m{1.5cm} m{2cm} m{0.1cm} m{1.8cm} m{1.2cm}}
&\underline{{\bf Thread 1} ($T_1$)} & & \underline{ {\bf Thread 2}  ($T_2$)} &\\

/* \underline{$T_{1}.a$} */ & {\tt lock()}; && &\\
& {\tt if} (flag == 0) && &\\
{\tt read}=$\{y\}$  & \hspace{3mm} x = ++y;&& &\\
{\tt write}=$\{x, y\}$  & {\tt else} && &\\
&\hspace{3mm}   x = (++y) + 5;&& &\\
  & {\tt unlock()};&& &\\
        &  &  $\searrow$ & & \\

&&  & {\tt lock()}; &  /* \underline{$T_{2}.a$ }*/\\
&&  & y = 2* x;    & {\tt read}=$\{x\}$  \\
&  &  & {\tt unlock()}; &  {\tt write}=$\{y\}$ \\

 &  &  $\swarrow$ & & \\

/* \underline{$T_{1}.b$} */ & {\tt lock()}; && &\\

{\tt read}=$\{y\}$  & y = y/2;&& &\\
{\tt write}=$\{ y\}$  & {\tt unlock()};  && &\\

\end{tabular}
}

\caption{ An example of shared-memory multithreading.}
\label{fig:simple-example}
\end{figure}

\section{System Model}
\label{sec:model}

Before we formally describe the provenance algorithm ($\S$\ref{sec:algorithms}),  we first present the system model assumed by \projecttitle ($\S$~\ref{sec:model}).

\myparagraph{Memory consistency model} Our approach relies on the use of the
Release Consistency memory model (RC)~\cite{DSM-RC}, which requires that all shared memory accesses are done via synchronization primitives.
This memory model  requires writes made by one thread to become visible only to another thread after the first thread releases a synchronization object and before the second thread acquires a synchronization object. 
For our purposes, this model has the critical
benefit of allowing us to restrict inter-thread communication (i.e. shared
memory accesses) to the synchronization points. By reducing the number of 
points in an execution where inter-thread communication can occur, we avoid
having to track individual {\tt load}/{\tt store} instructions, (given that any of these accesses could potentially be inter-thread communication), 
which would be extremely inefficient with current hardware. 

A natural consequence of the RC memory model is that the granularity of sub-computations for provenance in \projecttitle is the sequence of instructions between two \pthreads API calls instead of  individual {\tt load}/{\tt store} instructions. Note that the RC memory model is weaker
than, for example, the Sequential Consistency model (SC)~\cite{scLamport}, but
still guarantees correctness and liveness for applications that are data race
free. In fact,  the semantics provided by
\projecttitle is as restrictive as the POSIX specification~\cite{pthreads-spec}, which mandates that all accesses to shared data structures must be properly synchronized using 
\pthreads synchronization primitives. 

\myparagraph{Synchronization model} We support the full range of synchronization
primitives in the \pthreads API, including {\em mutexes}, {\em cond\_wait/cond\_signal}, {\em semaphores},  and {\em
barriers}. However, due to the weakly consistent RC memory model, our approach
does not support {\em ad-hoc synchronization mechanisms} such as user-defined spin locks. 
Although, ad-hoc synchronization mechanisms are used due to either flexibility
or performance reasons, in practice they are shown to be
error-prone introducing bugs or severe performance issues~\cite{adhoc-sync}.

\section{Design}
\label{sec:algorithms}
In this section, we first formally define the CPG ($\S$\ref{subsec:cpg}), and then present the algorithm to build the CPG ($\S$\ref{subsec:prov-algo}).

\subsection{Concurrent Provenance Graph} 
\label{subsec:cpg}

 We define the {\em
Concurrent Provenance Graph} (CPG)  as a directed acyclic graph $G =
(V,$ $E)$ with vertices ($V$) and edges ($E$). The
vertices of the CPG represent {\em sub-computations}. The edges represent the dependencies between the sub-computations. We distinguish between three kinds of edges: control, schedule, and data-dependence edges for recording control, schedule, and data dependencies, respectively. 

\myparagraph{Sub-computations} We define a {\em sub-computation}  as the sequence of instructions
executed by a thread between two \pthreads synchronization API calls. Using the RC memory model, we derive the synchronization and data dependencies at the granularity of a sub-computation.
We further divide each sub-computation as sequence of code thunks, or {\em thunks} to record the control path taken by the executing thread within the sub-computation.

\myparagraph{Dependencies} We distinguish between three kinds of dependencies: control, synchronization, and data dependencies. We next described these dependencies.

\myparagraph{I: Control edges} {\em Control edges}  are used to record the intra-thread causal order between sub-computations of the same thread
based on their execution order. Furthermore, we also record all control path taken by the executing thread within each sub-computation during the execution at the granularity of {thunks}.  

We model the execution of thread $t$ as a sequence of sub-computations ($L_t$).  Sub-computations in a thread are totally ordered based on their execution order using a monotonically increasing thunk counter ($\alpha$). We refer a sub-computation of thread $t$ using the counter $\alpha$ as an index in the thread execution sequence ($L_t$), i.e., $L_t[\alpha]$.  

We refer a thunk ($L_t[\alpha].\Delta$) as a sequence of instructions between two successive branches within each subcomputation. We denote a thunk of sub-computation $L_t[\alpha]$ using a counter $\beta$ as an index in the sub-computation as $L_t[\alpha].\Delta[\beta]$.

\myparagraph{II: Synchronization edges}  {\em Synchronization edges}   are
used to record the inter-thread causal order between sub-computations based on the synchronization order between threads. 
We derive synchronization edges based on the ordering of synchronization operations (also known as a {\em sync schedule}). In particular,  we build on the observation that synchronization primitives can be modeled as {\em acquire} and {\em release} operations. That is,   during synchronization, the synchronization object is {\em released} by one set of threads and subsequently  {\em acquired} by a corresponding set of threads blocked on the object~\cite{fast-track-pldi}. For example, an {\tt unlock(S)} operation releases $S$
and a corresponding {\tt lock(S)} operation acquires it. Likewise, a global {\tt barrier(S)} operation causes all threads to release $S$ as they reach the
barrier, and then causes them to re-acquire $S$. Similarly, all other
synchronization primitives can also be defined using acquire and release
operations~\cite{djit, fast-track-pldi}, which allows us discuss synchronization only in terms of release and acquire operations.

We derive the partial order based on the happens-before relation
($\rightarrow$)~\cite{djit,fast-track-pldi} between acquire and release operations. In particular, a release
operation happens-before the corresponding acquire operation.
Formally, two sub-computations $L_{(t_1)}[\alpha_1]$ and
$L_{(t_2)}[\alpha_2]$ are ordered by the happens-before relationship ($L_{(t_1)}[\alpha_1] \rightarrow
L_{(t_2)}[\alpha_2]$) if:  ($i$)  they are sub-computations of the
same thread ($t_1 = t_2$), and $L_{(t_1)}[\alpha_1]$ was executed before $L_{(t_2)}[\alpha_2]$; ($ii$)  $L_{(t_1)}[\alpha_1]$  is a release and $L_{(t_2)}[\alpha_2]$ is corresponding acquire on the same synchronization object $S$; ($iii$) due to transitivity if  
 $L_{(t_1)}[\alpha_1] \rightarrow L_{(t_3)}[\alpha_3] $ and $L_{(t_3)}[\alpha_3]  \rightarrow L_{(t_2)}[\alpha_2]$.

\myparagraph{III: Data-dependence edges}  {\em Data dependence edges}  records the flow of data between sub-computations of the same or different threads. 
 We derive the data dependencies between sub-computations using the read/write sets, and recorded partial order in control and synchronization edges. For a sub-computation $L_t[\alpha]$, the {\em read-set}
($L_t[\alpha].R$) and the {\em write-set} ($L_t[\alpha].W$) are the set of
addresses that were respectively read from and written to by
the thread while executing the sub-computation. 

Essentially, data dependence edges establish the {\em update-use relationship} between sub-computations. The update-use relationship exists between two sub-computations if they can be  ordered based on the happens-before relationship, and the write-set of the  precedent  sub-computations transitively intersects with the read-set of the antecedent sub-computations.

 \subsection{ Provenance Algorithm}
 \label{subsec:prov-algo}

At high-level, our algorithm records the multithreaded execution to construct the CPG.
Algorithm~\ref{fig:ithreadsRecord} presents the overview of the provenance algorithm, and details of the subroutines are presented in Algorithm~\ref{fig:ithreadsRecordProc}.
 
%
%
\begin{algorithm}[t]

\myfontsize
\SetLine
$\forall S, \forall i\in \{1, ...,T\} : C_S[i] \leftarrow 0$; // All sync clocks set to zero\\
\underline{{\bf  executeThread($t$)}}\\
\Begin{

	 {\tt initThread}($t$)\;
	\While{$t$ has not terminated}{
	
		{\tt startSub-computation}({\tt instruction})\;
		\Repeat{$t$ invokes synchronization primitive}{
			Execute {\tt instruction} of $t$\; 
			\If {({\tt instruction} is {\tt load} {\bf or} {\tt store})} {
			{\tt onMemoryAccess}({\tt instruction})\; 
			}
			\If {({\tt instruction} is {\tt branch} } {
			{\tt onBranchAccess}({\tt instruction})\; 
			}
		}
		$\alpha \leftarrow \alpha + 1$; // Increment sub-computation counter \\
		// Let $S$ denote invoked synchronization primitive\\
		{\tt onSynchronization}($S$)\;

	}
}

\caption{\bf  Data provenance algorithm}
\label{fig:ithreadsRecord}
\end{algorithm}

\myparagraph{Overview}   The provenance algorithm (shown in Algorithm~\ref{fig:ithreadsRecord}) is executed by all threads in parallel. During a thread execution, the thread traces memory accesses on {\tt load/store} instructions, and adds them to the read and
the write set of the executing sub-computation for deriving data dependencies. Additionally, the executing thread traces all branch instructions, and adds this information for thunks of the executing sub-computation to record control dependencies.
The thread continues to execute instructions until a synchronization primitive call is made to the \pthreads  library. At the synchronization point, we define the end point for the executing sub-computation. 
Thereafter, we let the thread perform the actual synchronization operation. At synchronization points, the algorithm derives control  and synchronization edges at the granularity of sub-computation by recording the happens-before order between sub-computations. Finally, we start a new sub-computation and repeat the process until the executing thread
terminates.

%
	\begin{algorithm}[t]
\myfontsize
\SetLine

\underline{{\bf initThread($t$)}}\\
\Begin{
$\alpha \leftarrow 0 $; // Initializes sub-computation counter ($\alpha$) to zero\\
$\forall i\in \{1, ...,T\}: C_t[i] \leftarrow 0$;  // $t$'s clock set to zero

}

\underline{{\bf startSub-computation({\tt instruction})}}\\

\Begin{

$\beta\leftarrow 0$; // Initialize thunk counter\\

$ L_t[\alpha].\Delta[{\beta}]\leftarrow$ {\tt instruction}; // Start new thunk

$C_t [t] \leftarrow \alpha$; // Update thread clock with sub-computation counter ($\alpha$) value \\

// Set sub-computation clock value to thread $t$'s clock
 
$\forall (i\in \{1, ...,T\}): L_t[\alpha] .C[i] \leftarrow C_t[i]$; 

			
}

\underline{{\bf onMemoryAccess({\tt instruction})}}\\ 
\Begin{
// Update read/write sets of the executing sub-computation \\

\uIf{{\tt instruction} is {\tt load} }{
$ L_t[\alpha].R \leftarrow L_t[\alpha] .R \cup \{${\tt pageID}$\}$; // On read access\\
}
\Else{
 $L_t[\alpha] .W \leftarrow L_t[\alpha] .W \cup \{${\tt pageID}$\}$; // On write access
}

}

\underline{{\bf onBranchAccess({\tt instruction})}}\\ 
\Begin{

$\beta \leftarrow \beta + 1$; // Increment thunk counter \\

$ L_t[\alpha].\Delta[\beta]\leftarrow$ {\tt instruction}; // Add a new thunk
}

%
%
%
%
\underline{{\bf onSynchronization($S$)}} \\
\Begin{
	
	\Switch{Syncronization type}{
		 \Case{{\tt release}($S$):}{
		 	// Update $S$'s clock to hold max of its and $t$'s clocks\\
	 		$\forall i\in \{1, ...,T\} :  C_S[i] \leftarrow max(C_S[i], C_t[i])$;\\
			sync($S$); // Perform the synchronization
		}
	 
	 	 \Case{{\tt acquire}($S$):}{
		 	sync($S$)\; 
			// Update $t$'s clock to hold max of its and $S$'s clocks\\
    			$\forall i\in \{1, ...,T\} :  C_t[i] \leftarrow max(C_S[i], C_t[i])$;
			
   		 }
        }
  }

\caption{\bf Subroutines for the provenance algorithm}
\label{fig:ithreadsRecordProc}
\end{algorithm}
	
%
%
%
%
%

\myparagraph{Details} For the CPG, control and synchronization dependencies are derived by 
happens-before ordering of sub-computations. To do so, we use vector clocks
($C$)~\cite{Mattern89virtualtime},  a widely used mechanism to generate a partial order of events and to infer causality. Our use of vector
clocks is motivated by its efficiency for recording a partial order between sub-computations in a complete decentralized manner instead of having to serialize all synchronization events in a total order.

In particular, each thread maintains a vector clock, i.e., an array/vector of size equal to the number of threads in the system.  
During a synchronization event, the clock of the thread performing the
acquire operation is updated based on the clock value of the thread performing
the release operation.  More precisely, the vector clock is updated as follows: if a thread $t_2$ acquires the synchronization
object $S$ released by a thread $t_1$, then each entry in $t_2$'s vector is
updated to hold the maximum of its old value and the corresponding value of
$t_1$'s vector at the moment of release.

To implement this mechanism, our algorithm maintains vector clocks for three kinds of
entities: threads,  synchronization objects, and sub-computations.  A {\em thread clock}
($C_t$) for a thread $t$ tracks the local logical time of the thread, which is
incremented each time a new thunk is created. A {\em synchronization clock}
($C_S$) for a synchronization object $S$ acts as a messaging medium  between
threads synchronizing on $S$ to update the thread clock. Finally, a {\em  sub-computation clock}
($L_t[\alpha].C$) determines the position of the sub-computation $L_t[\alpha]$
in the CPG, and is set to the clock value of the thread while executing the
sub-computation .

Based on the intuition developed so far, we next present the
subroutines used in the recording algorithm (see
Algorithm~\ref{fig:ithreadsRecordProc}). Let $T$ denote the number of threads in the system, which are numbered from $1$ to $T$.  Initially, each thread $t$ initializes (using routine {\tt initThread}($t$)) its monotonically increasing thunk
counter ($\alpha$) and the thread clock ($C_t$) to zero. In addition, vector clocks ($C_S$) of all synchronization objects $S$ are also initialized to zero. 
In the beginning of a new thunk (using routine {\tt startSub-computation()}), the clock value ($C_t$) of the thread $t$ is
updated based on the sub-computation  counter ($\alpha$) to keep track of the local logical time of $t$. The thread clock is updated by assigning the $\alpha$ to $t^{th}$ index of the thread clock $C_t[t]$. The updated value of thread clock ($C_t$)   is also assigned to the sub-computation's clock ($L_t[\alpha].C$). Finally, the read set and the write set
($L_t[T_t].R/W$) of the new sub-computation are initialized to empty set.

During a sub-computation execution, we trace reads and writes  (using routine {\tt onMemoryAccess()}) at the granularity of the
memory pages ({\tt pageID}), and update the respective read/write set
($L_t[T_t].R/W$) of the executing sub-computation.  

Similarly, we also trace branch instructions (using routine {\tt onBranchAccess()}), and update the thunk within the executing sub-computation.

At synchronization points, we define the end of the current sub-computation, and therefore, we increment the sub-computation  counter ($\alpha$) by one. The executing thread performs the synchronization operation (using routine {\tt onSynchronization()}). Recall that in our model, a synchronization operation is either a release or an acquire operation. Therefore, we handle {\tt onSynchronization()} accordingly. If it is a release operation on the synchronization object $S$ by the thread $t$, the 
releasing thread updates the synchronization object's clock ($C_S$) to
hold the maximum of its own clock value ($C_{t}$) and the clock  ($C_S$) of $S$. Then the releasing thread performs the actual release operation  on object $S$. Alternatively, if its an acquire operation then the acquiring thread first performs the acquire  operation on object $S$. After the acquire operation on the synchronization object $S$ by
thread $t$, the acquiring thread updates its own clock ($C_{t}$) to hold the
maximum of the clock value  ($C_S$) of $S$  and its own clock value ($C_{t}$).  
In this way, the synchronization clock ($C_S$) acts as a propagation medium to pass the
vector clock value from the thread doing the release to the thread doing the acquire operation.

In the end of the provenance algorithm, all sub-computations (along with their read/write sets) have a recorded value of sub-computation's vector clock ($L_t[\alpha].C$). The standard comparison of vector clocks defines the  happens-before partial order, through which causal order is derived between sub-computations.

\section{Implementation}
\label{sec:implementation}

\begin{figure}[t]

\centering
      \includegraphics[scale=.35]{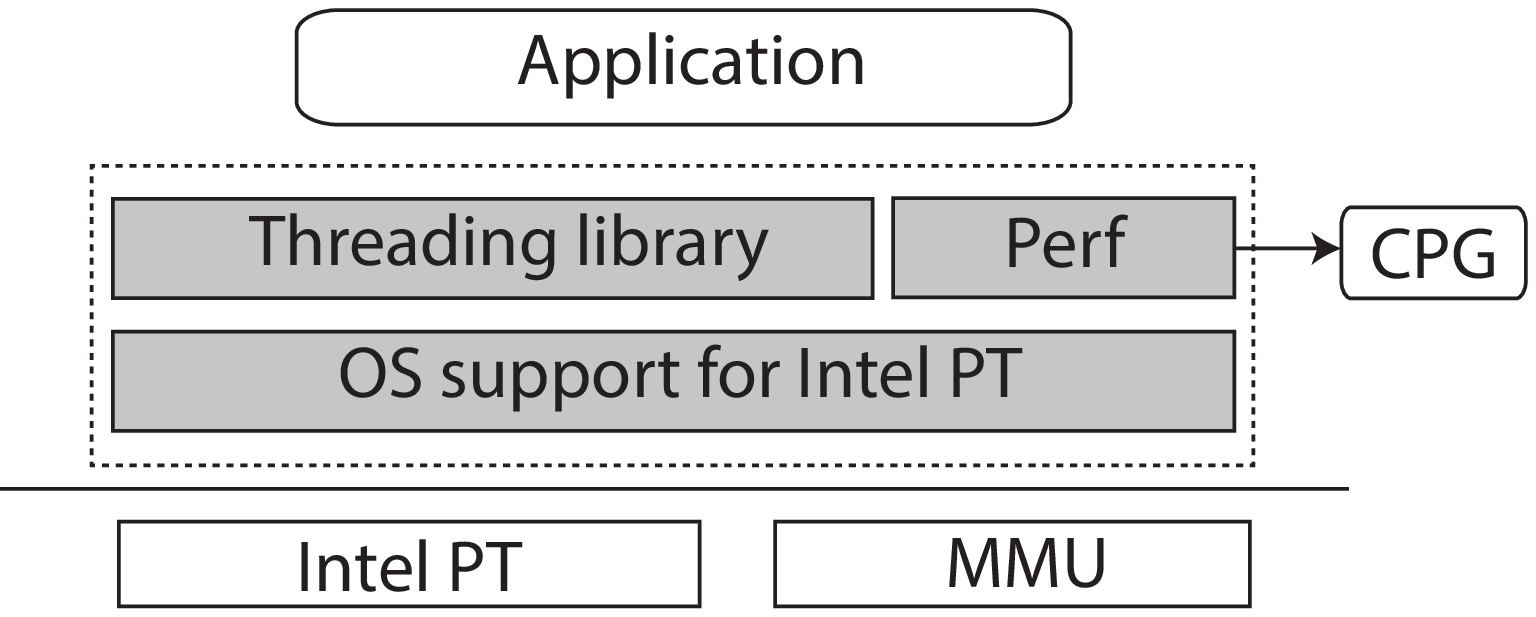}
  \caption{\projecttitle architecture.}
   
  \label{fig:basicSystem}

\end{figure}

This section describes the architecture and implementation of \projecttitle. We implemented \projecttitle as a dynamically linkable shared library for the GNU/Linux OS that can be loaded and linked  at runtime for {\tt POSIX} threads (replacing the \pthreads library). The application executables can simply link the library (without any recompilation) either using {\tt LD\_PRELOAD} or the {\tt -rdynamic} flag, specifying the path of the \projecttitle library.  The \projecttitle library exports the CPG as an extended interface in the {\tt perf} utility for supporting data provenance.   The architecture of \projecttitle (shown in Figure~\ref{fig:basicSystem}) consists of two main components: threading library ($\S$\ref{sec:impl-lib}) and OS support for \intelpt ($\S$\ref{sec:impl-OS}). We next describe these two components in detail.

\subsection{Threading Library}
\label{sec:impl-lib}
\begin{figure}[t]

\centering
      \includegraphics[scale=.35]{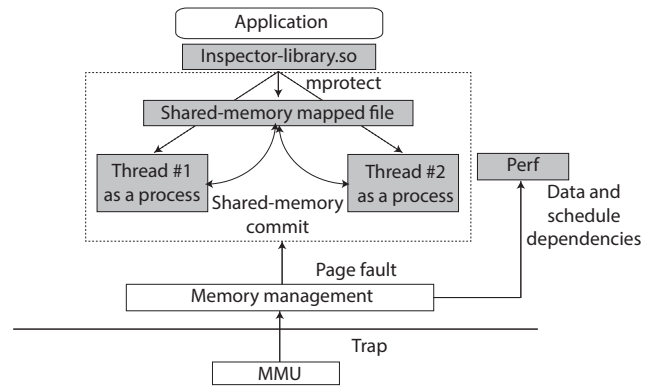}
  \caption{Architecture of the threading library.}
   
  \label{fig:lib-architecture}

\end{figure}

The threading library derives the data and schedule dependencies. The architecture of the threading library is shown in Figure~\ref{fig:lib-architecture}.

\myparagraph{Memory protection} A central challenge of the implementation of the
algorithm is keeping track of the data dependencies for shared-memory accesses
by all possible interleaving threads. Since monitoring every load and store to
each memory word would be too costly, we instead rely on the OS's
(hardware-assisted) segmentation fault mechanism to keep track of reads and writes at the granularity of memory pages.

To derive the read and write sets during sub-computation execution,  \projecttitle~uses standard memory protection  mechanism and signal handlers. In particular, \projecttitle protects the address space using {\tt mprotect(PROT\_NONE)} at the beginning of each sub-computation. This forces a trap (and the corresponding OS signal) the first time a page is read or written to in a given sub-computation. The respective signal handler, which is implemented by the \projecttitle library, records the information about the access, and also resets the protection bits so that subsequent accesses to the same page by the same thread in the same sub-computation can proceed without generating a trap. 

However, a naive page protection mechanism raises an important problem because all threads in a process share the same virtual memory
structures (namely the TLB and page table entries with the respective protection bits). This makes it difficult to keep track of which
threads are responsible for which memory accesses or to enforce different protections for different threads. Otherwise, we need to re-protect the page after serving every load and store instruction causing a large number of segmentation faults. 
To address this problem, \projecttitle implements threads as separate
processes (an idea  proposed by Grace~\cite{grace-oopsla-2009} and  Dthreads~\cite{dthreads-sosp-2011}).

\myparagraph{Threads as processes} \projecttitle  implements threads as separate processes thus allowing each thread has its own private address space and control over the virtual
memory structures.   This gives us the ability to manipulate the page protection of threads individually while providing a simple way to implement the release consistency memory
model. In particular, \projecttitle uses the {\tt clone} system call to fork off a new process on {\tt pthread\_create()}. The process that implements the newly created thread (i.e., the child process) already
shares parts of the execution context with the parent process (which implements the calling thread) such as file descriptors and signal handlers. 

But this raises a new problem, which is that, unlike threads, processes do not share their address spaces. We address this by taking advantage of the RC memory model we defined for \projecttitle, where threads share the updates only at synchronization points.  

\myparagraph{Shared memory commit} To implement the RC memory model, we use shared memory commit (originally proposed in distributed shared memory architectures such as TreadMarks~\cite{treadmark} and Munin~\cite{munin}) 
that allows threads to communicate at well-defined synchronization points.
Our shared memory commit is implemented using memory mapped files. In
particular, the virtual address ranges for the shared portions (globals and heap) of
the address space are mapped to memory mapped files, which are managed by the
\projecttitle library. These address ranges correspond to the heap and
the static (i.e., globals) regions.  During thread creation,
\projecttitle marks these address ranges as a private copy-on-write
mapping (using {\tt MAP\_PRIVATE} in {\tt mmap()}). The effect of this
is that whenever the child thread tries to write to a memory location,
the OS makes a thread-private copy of the memory page containing the
modification.  At synchronization points, the thread computes a {\em diff}
for each dirty page by performing a byte-level comparison between the
dirty page and the shared page. The deltas are then atomically
copied to the shared memory page; if there are overlapping writes
to the same memory location we resolve them using a last-writer wins policy.

\myparagraph{Input support} In addition to providing wrappers for {\tt pthreads} and {\tt malloc} related API calls, we also implemented shim layer for a number of input {\tt glibc} library calls  to record the data-flow from the input. For instance, we provide wrappers for {\tt mmap} for reading the input. In particular, the threading library differentiates between the {\tt mmap} calls made by the library itself and the target application. This allows us to record the mapping of the input file in the input address space. And, as described before, the library uses {\tt mprotect()} to derive the data flow from the input.

\subsection{OS Support for \intelpt}
\label{sec:impl-OS}
To obtain the control flow dependencies, we use Intel Processor Trace (PT) ISA extensions. We next present the implementation details of the OS support for \intelpt, recently released as part of the Broadwell (also in Skylake) micro-architecture. 

\myparagraph{Intel Processor Trace (Intel PT)}   \intelpt is an extension of Intel Architecture that logs information about software execution with minimal performance impact. The processor collects information such as control flow, execution modes and timings and formats it into highly compressed binary packets. Traditionally, Intel architectures provided Branch Trace Store (BTS) for tracing branch execution. However, BTS was slow and imprecise. Therefore, it was not adopted in practice. To overcome the limitations of BTS,  Intel recently introduced PT ISA extensions as part of the Broadwell  (also available in Skylake) micro-architecture.

\begin{figure}[t]

\centering
      \includegraphics[scale=.35]{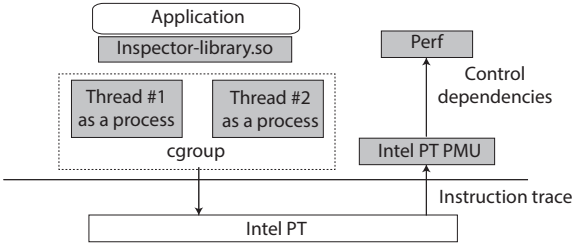}
  \caption{Architecture of the OS support for \intelpt.}
   
  \label{fig:pt-OS-support}

\end{figure}

\myparagraph{OS support} The \intelpt tracing facility is integrated into the operating system, which makes it possible to use  different trace buffers for different processes, and to make the facility available for non-root users.  In Linux this processor feature is exposed to the user-space as a Performance Measuring Unit (PMU) in the {\tt perf} event interface. We make use of the \intelpt PMU to derive the control flow dependencies. Figure~\ref{fig:pt-OS-support} shows the architecture for the OS support for \intelpt.

In particular, the {\tt perf} interface on Linux consists of a syscall, which gives back a file
descriptor. Events are accessed by obtaining buffers via {\tt mmap(2)} and can be
further controlled via {\tt ioctl()} syscall on the given file descriptor. Along with
interface the user-space {\tt perf} allows to dump and filter from these
buffers. In our case, this filtering is done by using Linux control
groups (also known as {\tt cgroups}). {\tt cgroups} is a kernel feature to apply constraint
like resource usage to a group of processes. It has the property, that by
default every child process belongs to the same process as its parent. Also for
{\tt perf\_events} such a {\tt cgroup} exists.

We create such a {\tt cgroup} exclusively for the application using \projecttitle. This is done because
our threading library causes applications using threads to create multiple
processes instead, whose process ids are not known in advance.

The subcommand {\tt perf record} is then used to dump the trace produced by \intelpt.
\intelpt generates a stream of TNT packets, which denotes the
conditional branches taken and TIP packets for indirect branches and function
returns. The data is referenced as a sample event in the {\tt perf} event list and
stored in a ring buffer called \emph{AUX area}. If perf tool cannot
keep up with processor trace it is possible (for example an interrupt occurs),
there will be gaps in the trace. 
(We provide a snapshot facility ($\S$\ref{sec:snapshot}) to overcome this limitation.)

After execution the result can be further processed by using a set of tools
for example {\tt perf script}. The branch information is still in a compressed
form and needs to be decoded. We make use of the Intel Processor Decoder Library for Intel PT  that is integrated in the {\tt perf} utility. 
To map the trace onto binaries, it needs access to executables and linked
libraries of the application. For that, we track mmap events to
know the location of each loadable during the execution.

Along with \intelpt, the page fault events generated by the kernel will be included
additionally to the trace packets. Because our threading library uses {\tt mprotect}
syscall to monitor access of heap and global memory space, whenever the
application access this memory, the MMU will generate a page fault. These
page faults also include the location, where in code memory was accessed.

\section{Snapshot Mechanism}
\label{sec:snapshot}
An additional challenge that we need to address in the implementation of \projecttitle is to deal with the excessive log data produced by \intelpt, especially for long running applications. Therefore, we further extend the library to support a live snapshot facility, where the user (or an application using \projecttitle) can analyze the provenance on-the-fly while the program is still running. Thus, the snapshot facility provides a practical alternative to restrict the space overheads imposed for storing the CPG. 

For the snapshot facility, the library periodically takes a consistent cut of the CPG. A cut is {\em consistent} if, for any synchronization operation on object $S$ in the trace,  {\em acquire}($S$) operation being in the cut implies that corresponding {\em release}($S$) is also included in the cut~\cite{chandy-lamport}.  To achieve so, we make use of modeling synchronization primitives as {\em acquire} and {\em release} operations (described in $\S$\ref{sec:algorithms}). Each thread invokes the snapshot operation on the latest synchronization event ({\em acquire} or {\em release}) in the recorded trace.

We implemented the consistent cut facility using \intelpt interface for {\tt perf},
which provides mechanism for the full trace, and a snapshot mode.
When the full trace is enabled then the kernel does not overwrite the data that the user-space has not collected yet. 
This results in gaps in the trace, if the user-space process is not fast enough in collecting the log data. 
Whereas, in the snapshot mode, however, the old data in this ring buffer is constantly overwritten so that an application
can start and stop tracing around a certain event. 
The {\tt perf} tool exposes this feature by installing a handler on
signal {\tt SIGUSR2}, which triggers the start of a trace. \projecttitle makes use of the signal
and forwards it to {\tt perf} to record a consistent snapshot of the trace based on the aforementioned checkpointing mechanism.  Using this signal, we implemented a simple ring buffer with a configurable number of slots (each slot size is set to $4$MB). As the user (or the application using \projecttitle) finishes the live analysis on the recorded snapshots of the CPG, we reuse those slots for storing the new incoming snapshots of the CPG.

\section{Evaluation}
\label{sec:evaluation}
In this section, we present an experimental evaluation of \projecttitle based on the implementation described in  $\S$\ref{sec:implementation}. 
Our evaluation answers the following questions.

\begin{figure*}[htp]
  \subfloat[Time overheads]{\label{fig:overheads-time} \includegraphics[scale=0.28]{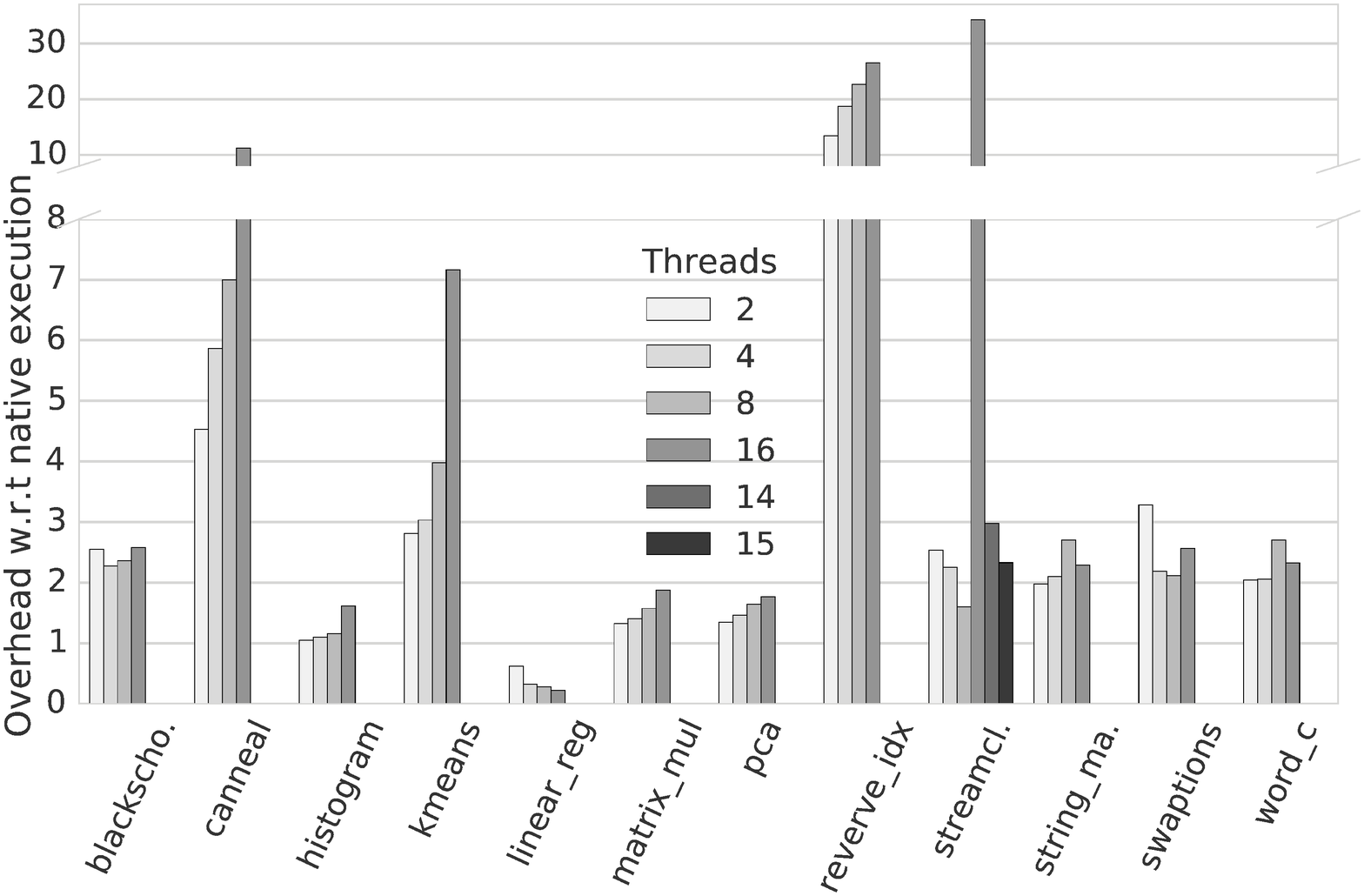}}
  \subfloat[Work overheads]{\label{fig:overheads-work} \includegraphics[scale=0.28]{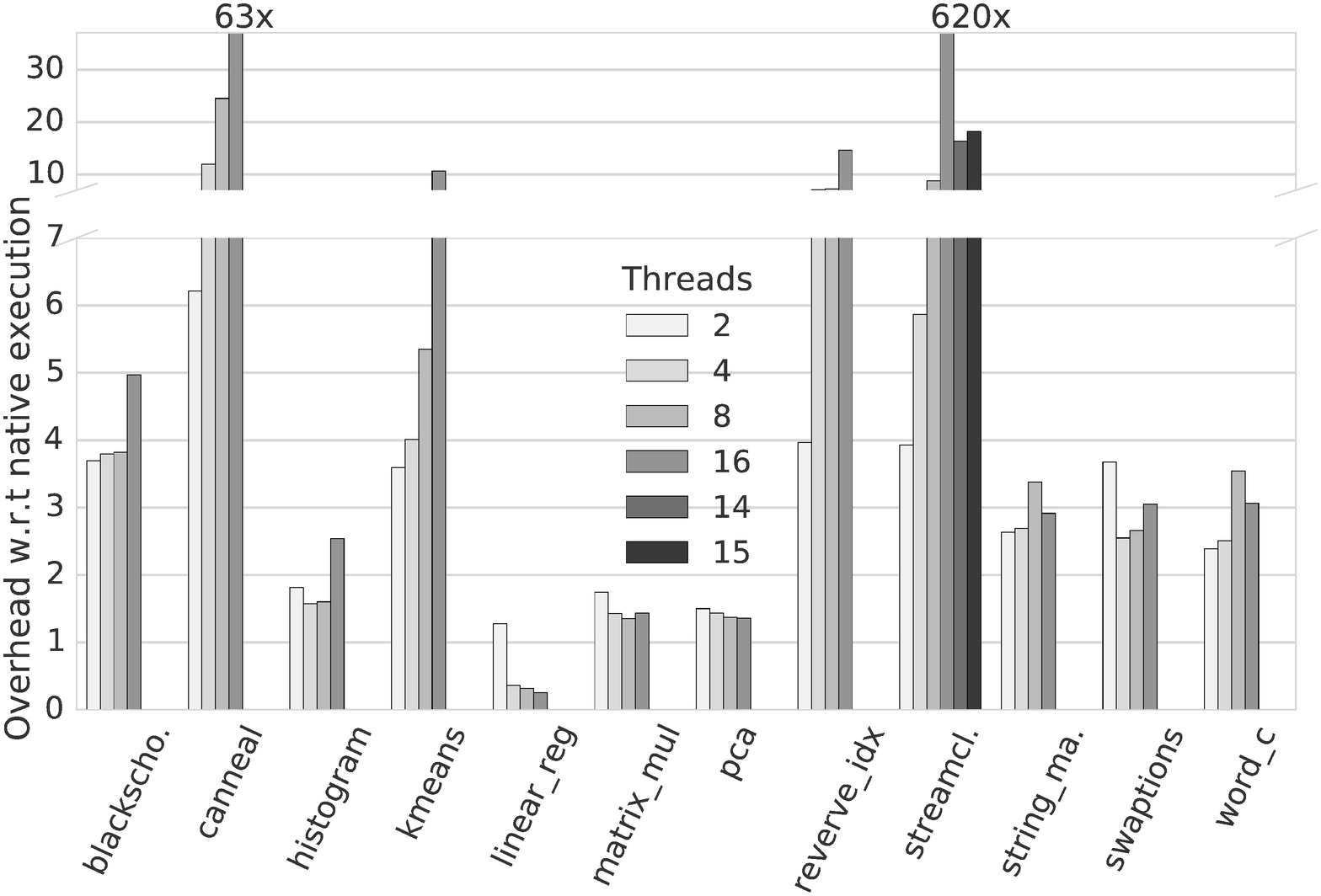}}

  \caption{\label{fig:overheads} Performance overheads of \projecttitle{} over native execution with increasing number of threads.}

\end{figure*}

\begin{itemize}
\item What performance overheads does \projecttitle impose for recording the provenance graph? ($\S$\ref{subsec:overheads})
\item What are the sources for these overheads? ($\S$\ref{subsec:performance-overheads-breakdown})
\item How do these overheads scale with increase in the size of the input data? ($\S$\ref{subsec:data-sizes-overheads})
\item What are the space overheads for the CPG? ($\S$\ref{subsec:overheads-breakdown})
\end{itemize}

\subsection{Experimental Setup}

\myparagraph{Experimental platform} We used an Intel Xeon processor based on Broadwell micro-architecture
as our host machine. The
host system consists of $8$ cores ($16$ hyper-threads) of Intel(R) Xeon(R) CPU Processor D-$1540$
($12$M Cache, $2.00$ GHz) and $32$ GB of DRAM main memory. The host
machine is running Linux with kernel $4.3.0$ in $64$-bit mode. 

\myparagraph{Applications and dataset}  We evaluated \projecttitle using applications from two multithreaded benchmark suites: Phoenix 2.0 \cite{phoenix} and PARSEC 3.0 \cite{parsec}. Table~\ref{tab:apps} lists the applications used for the evaluation along with the input data and benchmark parameters.

\myparagraph{Performance metrics: Time and Work}  For each run, we consider two types of measures: \emph{time} and
 {\em work}.  Time refers to the amount of (end-to-end)
run-time to complete the parallel computation.  Work refers to the total amount of
computation performed by all threads and is measured as the overall CPUs utilization for all threads. 
Both metrics are important
and complementary: time measurements reflect the end user perceived latency,
whereas work measurements assess the overall resource (CPU) utilization.

\myparagraph{Measurements} All applications were compiled using
GCC $5.2.1$ compiler with {\tt -o3} optimization flag. For all
measurements, we report the average over $6$ runs with minimum and maximum values
discarded (truncated mean).

 We measured work and time numbers for both \pthreads and \projecttitle executions with the same number of threads. For time measurements, we report the run-time comparison between the native {\tt pthreads} execution, and \projecttitle execution.   To measure work, we used the CPU accounting controller in {\tt cgroups} to account the CPU usage of all threads. 

Finally, the log produced by
{\tt perf} was written to {\tt /tmp} on {\tt tmpfs} to allow high throughput.
 
\begin{figure*}[htp]
  \subfloat[Time overhead breakdown]{\label{fig:overheads-breakdown-time} \includegraphics[scale=0.28]{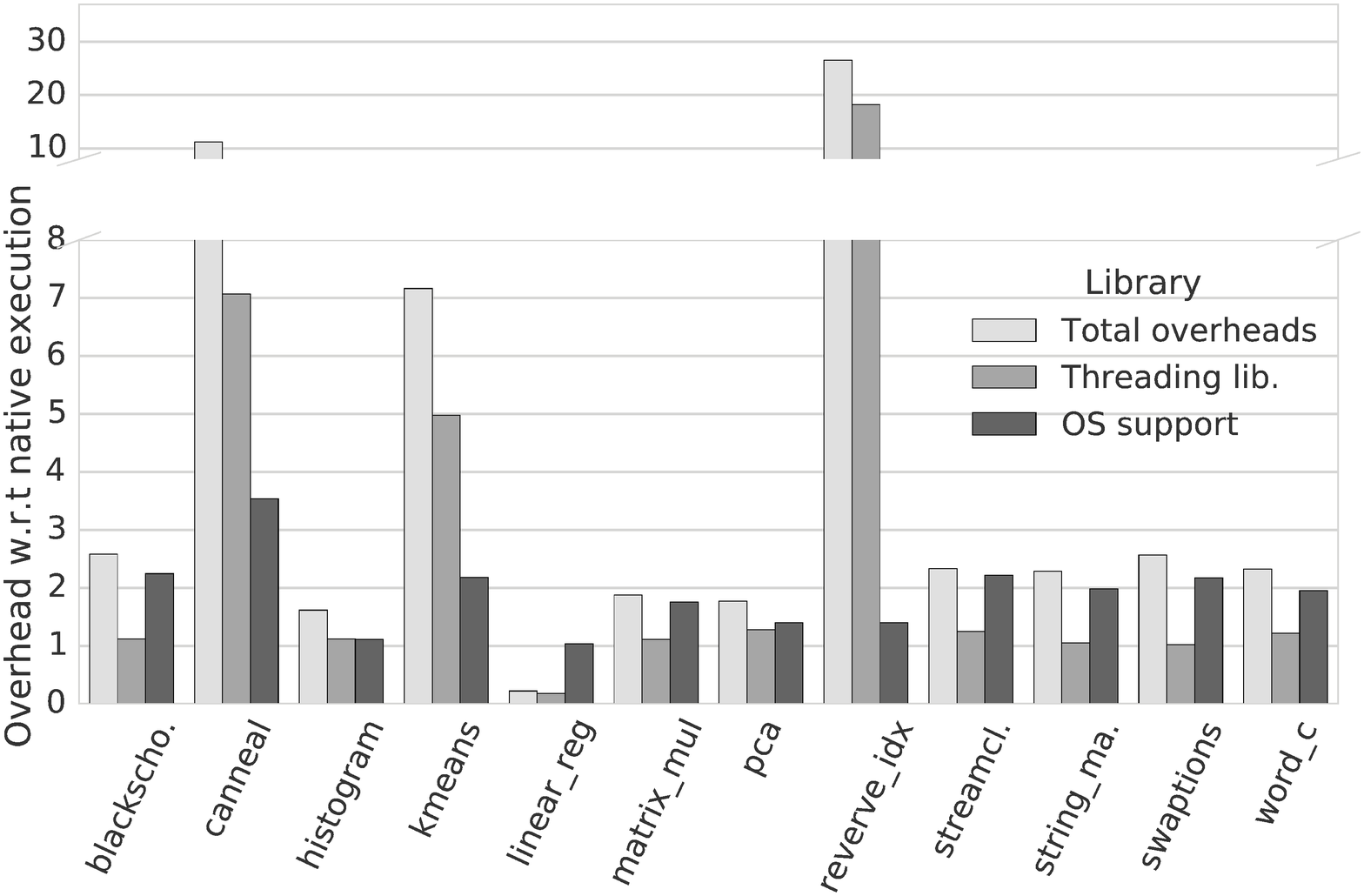}}
  \subfloat[Work overhead breakdown]{\label{fig:overheads-breakdown-work}
  \includegraphics[scale=0.28]{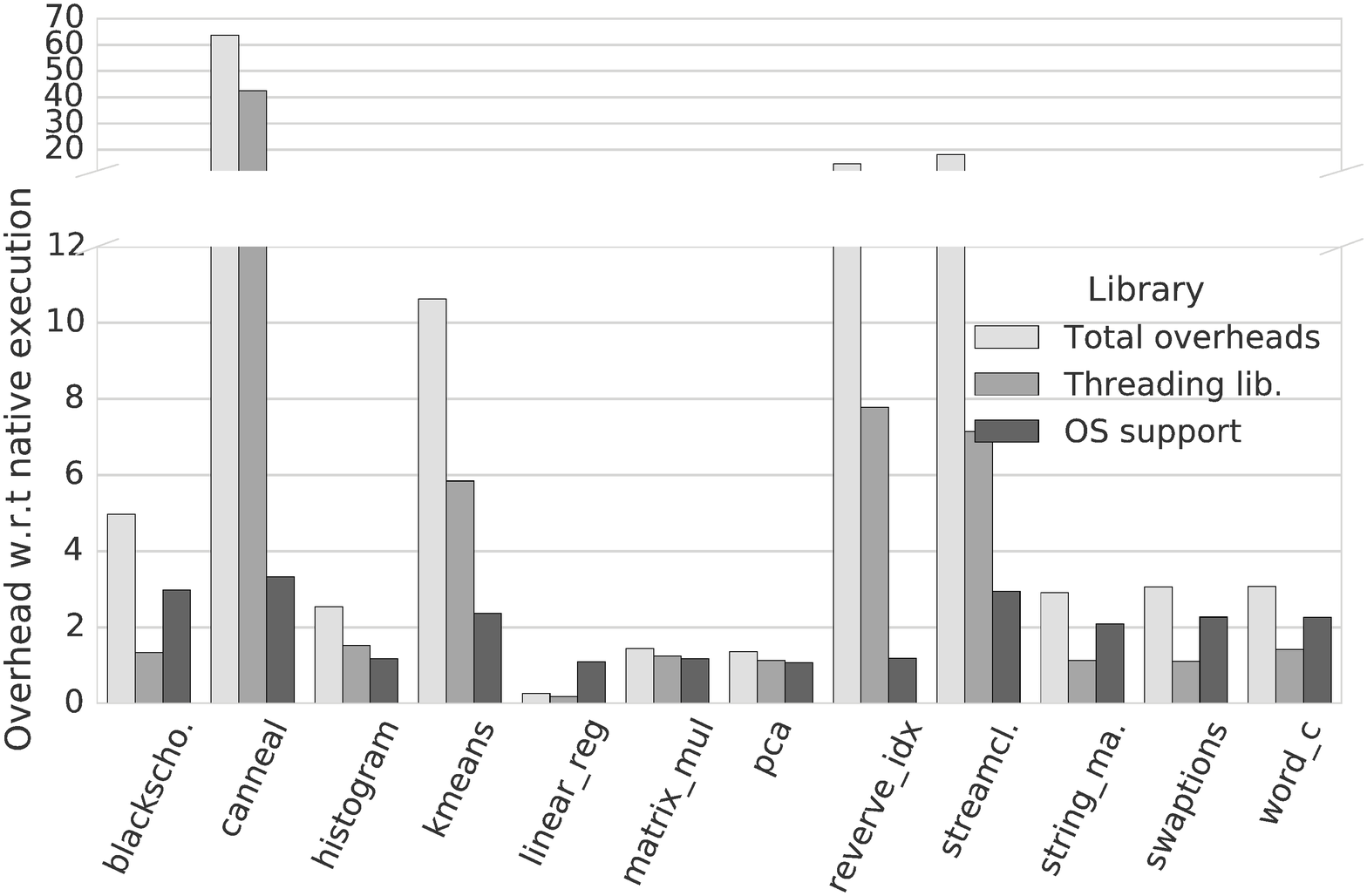}}

  \caption{\label{fig:overheads-breakdown} Performance overhead breakdown with $16$ threads --- except for {\em streamcluster}, where the breakdown for $15$ threads is shown.}

\end{figure*}

\subsection{Performance Overheads for Data Provenance}
\label{subsec:overheads}
First, we explain the provenance overheads imposed by \projecttitle w.r.t. the native \pthreads execution. Figure~\ref{fig:overheads} shows the provenance overheads of \projecttitle w.r.t. the native \pthreads execution with varying number of
threads (from $2$ to $16$ threads). As expected, the provenance overheads increases with the increase in the number of threads. This is because the shared memory commit ($\S$\ref{sec:impl-lib}) takes longer time with a higher number of threads, as each thread spends less time computing on the input data.

The experiment shows that the provenance overheads using \projecttitle vary across applications. 
We observe that a majority of applications ($9$/$12$) have a reasonable overhead between $1\times$ up to $2.5\times$ w.r.t. the native execution. However, three applications have exceptionally high overheads:  {\em canneal}, {\em reverse\_index}, and {\em kmeans}. The high overheads is explained as follows: {\em canneal} modifies a lot of memory pages that leads to a high number of page faults for deriving read and write sets (see Table~\ref{tab:apps}). Whereas, {\em reverse\_index} does a lot of small memory allocations across threads leading to a large number of segmentation faults. 
Finally, {\em kmeans} creates more than $400$ threads until the clusters co-efficient converges, when we specify $500$ as the parameter for the iterative convergence algorithm (see Table~\ref{tab:apps}). Since, creating a process takes more time than creating a thread, we see a slowdown in {\em kmeans}.

On the other hand, {\em linear\_regression} performs better than \pthreads, which is explained by the fact that our implementation of threads as processes ($\S$\ref{sec:implementation})  avoids false sharing, as previously noted by Sheriff~\cite{false-sharing-sheriff}, which leads to improved performance.

Lastly, in the case of {\em streamcluster}, we were limited by our physical memory to store the log in {\tt tmpfs} for $16$ threads (see $\S$\ref{subsec:overheads-breakdown}). Therefore, we also show the overheads with $14$ and $15$ threads,  where the provenance log can fit into the main memory.  To better understand the breakdown of provenance, we chose $15$ threads for {\em streamcluster} in $\S$\ref{subsec:performance-overheads-breakdown}.

\subsection{Performance Overheads Breakdown} 
\label{subsec:performance-overheads-breakdown}

Next, we investigated the breakdown of the provenance overheads. Recall that our system implementation has two major components: (1) the threading library ($\S$\ref{sec:impl-lib}), and (2) the OS support for \intelpt ($\S$\ref{sec:impl-OS}). 
Figure~\ref{fig:overheads-breakdown} shows the breakdown of overheads with $16$ threads normalized to the native \pthreads execution. We quantify the breakdown as the time taken by the threading library  and the OS support for \intelpt  . The result shows an interesting pattern: the applications with unreasonably high overheads ({\em canneal}, {\em reverse\_index}, and {\em kmeans}) spend a majority of time in the threading library for the above mentioned reasons. Whereas, the overheads for tracing the control flow due to \intelpt  is a dominant factor for the other applications. These results highlight that for a majority of applications ($9/12$) the underlying hardware is still a bottleneck to achieve low provenance overheads.

\subsection{Scalability with the Input Data}
\label{subsec:data-sizes-overheads}

\begin{figure*}[t]
\centering
{
\begin{tabular}{m{3.6cm}|m{5.2cm}| m{2.2cm}|m{2.3cm}}
   { Application} & Dataset / Parameters & Page faults & Faults/sec\\
  \hline \hline
    blackscholes& 16 in\_64K.txt prices.txt & 2.49E+04& 2.58E+04 \\
    canneal& 15 10000 2000 100000.nets 32 & 2.11E+06 & 21.57E+04 \\
    histogram& large.bmp & 4.27E+04 & 10.78E+04	  \\
    kmeans& -d 3 -c 500 -p 50000 -s 500 & 1.16E+06 & 13.99E+04  \\
    linear\_regression& key\_file\_500MB.txt & 2.88E+04 & 11.11E+04  \\
    matrix\_multiply& 2000 2000 & 2.32E+05 & 11.65E+04  \\
    pca& -r 4000 -c 4000 -s 100 & 5.34E+05 & 10.22E+04\\
    reverse\_index & datafiles & 2.61E+04 & 10.35E+04  \\
    streamcluster& 2 5 1 10 10 5 none output.txt 16 & 1.64E+05 & 1.163E+04\\
    string\_match &key\_file\_500MB.txt & 3.11E+04 & 1.993E+04\\
    swaptions & -ns 128 -sm 50000 -nt 16  & 4.66E+04 & 1.207E+04 \\
    word\_count& word\_100MB.txt	 & 1.56E+05 & 54.34E+04  \\

\hline
\end{tabular}
}

\caption{\label{tab:apps} Runtime statistics for all benchmarks with 16 threads}

\end{figure*}

 In addition to scalability w.r.t.  threads, we also measured the performance overheads with increase in the size of the input data.  For that, we
report the performance overheads for four applications that are available with
three input sizes: small ($S$), medium ($M$), and large ($L$). These four applications are: {\em histogram}, {\em linear\_regression}, {\em string\_match}, and {\em word\_count}.

\begin{figure*}[htp]
  \subfloat[Time overhead ]{\label{fig:data-size-overheads-time} \includegraphics[scale=0.28]{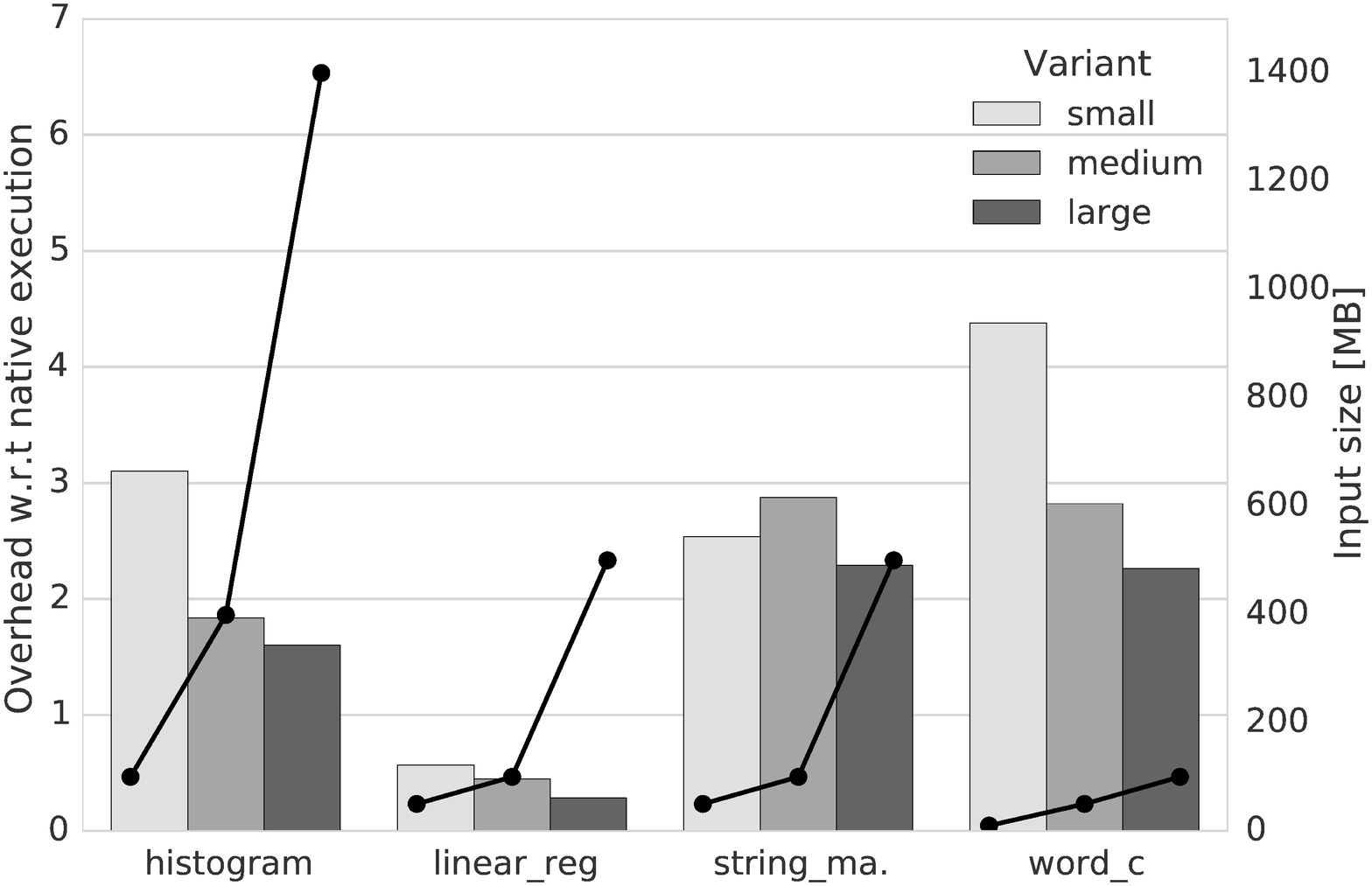}}
  \subfloat[Work overhead]{\label{fig:data-size-overheads-work}\includegraphics[scale=0.28]{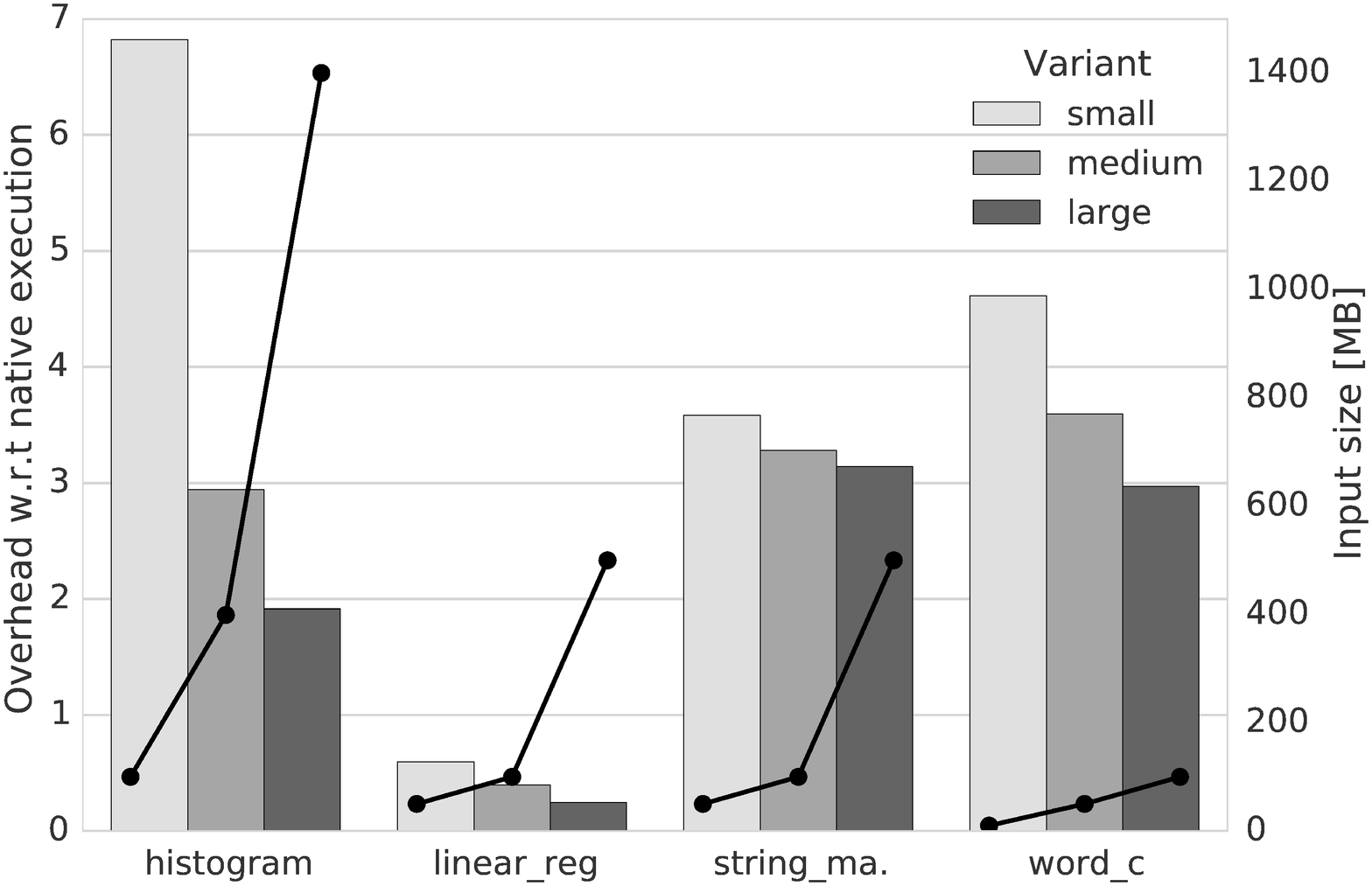}}

  \caption{\label{fig:data-size-overheads} Scalability of overheads with increase in the input data sizes with $16$ threads. }

\end{figure*}

In this experiment, we kept the number of threads to a constant  ($16$ threads), and we varied the input sizes for these applications.  Figure~\ref{fig:data-size-overheads} shows the results for our experiment. The bar plot shows the performance overheads w.r.t. to the native \pthreads execution on the $Y1$-axis for three input sizes ($S$, $M$, $L$). For the reference, the input sizes are also shown by a line plot in the same figure on the $Y2$-axis. 

The result shows that the gap between \pthreads and \projecttitle narrows with bigger input sizes. This is due to the fact that most applications use a data-parallel programming design pattern for parallelization, where the main threads divides the input data evenly between the worker threads. As the input size increases, each thread needs to perform more work (or compute on a larger input size) than the time spent for synchronization. As a result, each thread spends relatively more time outside the shared-memory commit to compute on the data, and thus, it results in improved performance.

\subsection{Space Overheads for the Provenance Graph}
\label{subsec:overheads-breakdown}

Finally, we present the space overhead for storing the provenance graph. A major limitation of using \intelpt is that it produces a large amounts of trace data. Furthermore, the threading library also produces trace data to record the data and schedule dependencies.  Table~\ref{tab:space-overheads} shows the space overhead for all applications with~$16$ threads. 

The space overheads vary across applications: it can be as low as $183$MB for {\em linear\_regression} and as high as $29.3$GB for {\em streamcluster}. The result shows a strong correlation between the log bandwidth and branch instructions with a correlation coefficient of 0.89, which was expected, because the
log consists of taken branches.

Fortunately,  the provenance log written by {\tt perf} turns out to be highly compressible. We
were able to achieve a compression ratio of between $6\times$ and $37\times$ times using the lz4 compression algorithm. 
Furthermore, the snapshot facility (described in $\S$\ref{sec:snapshot}) restricts the active area of space usage, and the user can reuse the space in the ring buffer after analyzing (or collecting) the provenance graph.

\section{Discussion: Case-studies}
\label{sec:discussion}
While data provenance is useful across a wide range of workflows, we discuss three active projects where \projecttitle is being used to increase the dependability, security, and efficiency of software systems. 

\myparagraph{Dependability: Debugging programs~\cite{fast-track-pldi}} Multithreaded programs are notoriously difficult to debug because of the inherent non-deterministic thread scheduling by the OS.  Currently, debugging techniques rely on examining the memory state during the program execution or by analyzing core dumps after the crash. These techniques mainly target ``what" is the state of the program without revealing much about ``why" is it the state of the program is like that. Our library can be extended to aid the developers to better understand the failed execution by augmenting the existing debugging techniques with the provenance of the memory state.

\myparagraph{Security: Dynamic Information Flow Tracking (DIFT)~\cite{dift}} DIFT protects software against data leaks by restricting the suspicious I/O calls. Our library can be extended to support DIFT by carrying a taint for the sensitive data as part of the provenance, and  restricting the output activities at the level of system calls. In particular, a policy checker can analyze the taint provenance to disallow sensitive data leaks. The policy checker can be embedded at the level of {\tt glibc} wrappers for the output system calls. Note that we currently target accidental or buggy, but not a malicious threat model because our library is a user-space solution.

\myparagraph{Efficiency: Memory management for NUMA~\cite{memprof}} The recent advancement in NUMA architectures offers a wide range of configurations for the interconnects with varying memory bandwidth, and  it is unclear how these different configurations affect the OS support for memory management. Our library can be extended to investigate the potential impact of interconnect topologies on memory management, and can be extended to  optimize the memory layout for a given interconnect topology. This optimization requires the memory access patterns that could be easily derived from the CPG.

\begin{figure*}[t]
\centering
{
\begin{tabular}{m{2.6cm}|m{1.8cm}|m{2.1cm}|m{1.6cm}|m{2.1cm}|m{2.3cm}}
       & \multicolumn{3}{c|}{ Provenance log details [MB] }   &  Bandwidth & Branch instr. \\
   { Application} & Size & Compressed & Ratio & [MB/sec] &  [Instr/sec] \\
  \hline \hline
    blackscholes& 851& 57.3 &15$\times$& 882& 2.49E+09 \\
    canneal& 5343& 315.0 & 17$\times$& 547& 1.55E+09 \\
    histogram& 381& 11.3 & 34$\times$& 961& 4.17E+09 \\
    kmeans& 11900& 522.0 &23$\times$& 1438& 5.79E+09 \\
    linear\_regression& 183& 5.5 &34$\times$& 707& 3.81E+09 \\
    matrix\_multiply& 2101& 97.0 &22$\times$& 105& 4.05E+08 \\
    pca& 1900& 116.0 &16$\times$& 364& 1.42E+09 \\
    reverse\_index& 192& 5.7 & 34$\times$& 764& 2.87E+09 \\
    streamcluster& 29300& 787.0 &37$\times$& 2083& 7.78E+09 \\
    string\_match& 2751& 430.0 &6$\times$& 1763& 5.61E+09 \\
    swaptions& 7061& 929.0 &8$\times$& 1830& 4.84E+09 \\
    word\_count& 4121& 508.0 & 8$\times$& 1435& 2.80E+09 \\

\hline
\end{tabular}
}

\caption{\label{tab:space-overheads} Space overheads for all benchmarks with 16 threads.}

\end{figure*}

\section{Related Work}
\label{sec:related}

Data provenance is a well-studied concept because of it's wide applicability in different complex computer systems. Next, we review the related work from different domains.

\myparagraph{Database systems} Provenance has been shown to be important in databases for materialized views, probabilistic databases, data integration, and curated databases (see a survey paper for more details~\cite{provenance-database-tutorial}). Almost, all existing provenance work in databases leverages the explicit database schema and structured layout of the input records in tables to build the provenance graph, whereas, \projecttitle does not assume any structured layout of the input data.

\myparagraph{``Big Data" analytics} Data provenance is being increasingly used in ``big data"  processing  for  debugging complex workflows~\cite{nova, conductor-podc-2010, conductor-ladis-2010, conductor-nsdi-2012}, and also for incremental computation~\cite{ Bhatotia15, slider, contraction-tree, shredder, incoop-hotcloud, incoop,   incApprox}.  These ``big data" systems leverage the underlying task-based programming model such as MapReduce~\cite{mapreduce} or Dryad~\cite{dryad} 
for building the provenance graph. 
In particular, these systems construct the provenance graph based on the data-flow graph generated from the data-parallel programming model. 
Instead of relying on the constrained task-based programming model,  \projecttitle derives the graph automatically for shared-memory multithreaded programs.

\myparagraph{Distributed and network systems} Many distributed and network systems propose provenance techniques for tracing the  execution of distributed protocols to provide accountability, fault detection, forensics, verifiability, network debugging, negative provenance~\cite{ wu-2014-negative-provenance, snp, dtap, wu-2014-negative-provenance}. To make the lineage secure in the presence of adversaries in distributed settings, they further embed techniques like tamper-proof logging~\cite{peer-review} along lineage for non-repudiability.  
These systems leverage the semantics of distributed protocols to derive a state-machine, and capture the lineage information by manually modifying the state-machine. Instead, we do not require any protocol-specific state-machine. Albeit, we currently do not support distributed systems.

\myparagraph{Storage systems} Storage systems supporting provenance collect meta-data of newly created objects in the system (via the OS support), 
and  maintain their lineage information such as the chain of ownership and the transformations performed on objects. In this context, one of the most important line of work is Provenance-Aware Storage Systems (PASS) that automates collection and maintenance of provenance~\cite{pass-atc} of objects in the system. 
 In addition, PASS also supports queries, tracing the lineage of objects, upon the provenance data. 
In contrast to PASS that tracks objects in storage systems, our focus is on tracing the lineage of shared-memory accesses in multithreaded programs at the granularity of memory pages. Like PASS, we also rely on the OS support for tracking of memory pages.

\myparagraph{Memory tracing} Our approach is complementary to numerous run-time~\cite{memtrace} and compile-time~\cite{cgo-compiler-provenance} tools that allow fine-grained byte-level memory read and writes made by threads. In contrast, our tool makes a trade-off of memory tracking at the granularity of memory pages, and uses a combination of OS support and the new ISA extensions to track the data flow for the entire program. Furthermore, we also record control and schedule dependencies for the multithreaded execution as part of the data provenance graph.

\myparagraph{Operating systems} Linux Provenance Module (LPM)~\cite{lpm} provides OS support to collect system-wide provenance. In contrast to LPM, \projecttitle is a user-space solution and does not require any modifications to the underlying OS. Secondly, unlike LPM, which collects provenance at the granularity of a process, we collect data provenance at a finer granularity of a thread.  On the other hand, LPM benefits from the integrated OS approach to secure the provenance information. 

\myparagraph{Programming languages} Programming languages researchers develop language-based provenance approaches relying on a new language with special data-types. These language-based approaches derive the provenance graph using techniques such as self-adjusting computation~\cite{Acar05}. In contrast, our work supports existing programs without relying on any language-level support or a new type system.

\section{Conclusion}
\label{sec:conclusion}

In this paper, we presented \projecttitle, a data provenance library for multithreaded programs. Our approach targets existing executables, relies on OS-specific mechanisms and new ISA extensions of \intelpt  to efficiently build the {\em Concurrent Provenance Graph (CPG)}. The CPG records control, data, and schedule dependencies for the shared-memory multithreaded program execution. Our solution is straightforward to deploy: it simply replaces the {\tt pthreads} library, allowing existing applications to benefit from our approach with no re-compilation or code changes. \projecttitle's source code is publicly available for further use in a wide-range of workflows for data provenance. The original paper of  \projecttitle is available onine~\cite{inspector-icdcs}.

\balance
\bibliographystyle{abbrv}

\bibliography{main}

\end{document}